\documentclass[prx,twocolumn,english,superscriptaddress,longbibliography]{revtex4-2}
\pdfoutput=1
\usepackage[unicode=true, breaklinks=false, backref=false, colorlinks=true, linkcolor=blue, urlcolor=blue, citecolor=blue]{hyperref}

\usepackage{latexsym,amsmath,amsthm,amssymb,amsfonts,bm,bbold,amstext,bbm,bm}
\usepackage{graphicx,relsize,dsfont,times,enumerate,float,cancel}
\usepackage{lipsum}
\usepackage[table]{xcolor}
\usepackage[utf8]{inputenc}
\usepackage[english]{babel}
\usepackage{cancel}
\usepackage{tcolorbox}
\definecolor{light_blue}{HTML}{f0f5ff}
\tcbset{fontupper=\normalsize,
colback=white, colframe=black, colbacktitle=light_blue,
coltitle= black, arc=0.5mm, center title}

\definecolor{comment}{HTML}{9326ff}

\renewcommand{\paragraph}[1]{\noindent\textbf{#1}---}

\usepackage[normalem]{ulem}
\newcommand{\stkout}[1]{\ifmmode\text{\sout{\ensuremath{#1}}}\else\sout{#1}\fi}
\newcommand{\tr}{\mathrm{Tr}}

\def\bra#1{\langle{#1}|}
\def\ket#1{|{#1}\rangle}
\def\braket#1{\langle{#1}\rangle}
\newcommand{\ketbra}[2]{\ket{#1}\!\bra{#2}}

\def\BraVert{\egroup\,\mid\,\bgroup}

\newcommand{\phantomdagger}{\phantom{\dagger}}

\newcommand{\TT}{\mathrm{TT}}
\newcommand{\singlet}[1][\,]{^1(\mathrm{TT})_{#1}}
\newcommand{\triplet}[1][\,]{^3(\mathrm{TT})_{#1}}
\newcommand{\quintet}[1][\,]{^5(\mathrm{TT})_{#1}}

\newcommand{\mytitle}{Quintet formation and exchange fluctuations: The role of stochastic resonance in singlet fission}

\begin{document}
\sloppy

\title{\mytitle}

\author{Miles I. Collins}
\email{miles.collins@unsw.edu.au}
\affiliation{\textit{School of Physics and ARC Centre of Excellence in Exciton Science, UNSW Sydney, NSW 2052, Australia}}

\author{Francesco Campaioli}
\email{francesco.campaioli@rmit.edu.au}
\affiliation{\textit{Chemical and Quantum Physics, and ARC Centre of Excellence in Exciton
Science, School of Science, RMIT University, Melbourne 3000, Australia}}

\author{Murad J. Y. Tayebjee}
\email{m.tayebjee@unsw.edu.au}
\affiliation{\textit{School of Photovoltaic and Renewable Energy Engineering, UNSW Sydney, NSW 2052, Australia}}

\author{Jared H. Cole}
\email{jared.cole@rmit.edu.au}
\affiliation{\textit{Chemical and Quantum Physics, and ARC Centre of Excellence in Exciton Science, School of Science, RMIT University, Melbourne 3000, Australia}}

\author{Dane R. McCamey}
\email{dane.mccamey@unsw.edu.au}
\affiliation{\textit{School of Physics and ARC Centre of Excellence in Exciton Science, UNSW Sydney, NSW 2052, Australia}}

\date{\today}

\begin{abstract}
\noindent Singlet fission describes the spin-conserving production of two triplet excitons from one singlet exciton. The existence of a spin-2 (quintet) triplet-pair state as a product of singlet fission is well established in the literature, and control of quintet formation is an important step towards applying singlet fission in photovoltaics and quantum information. However, a definitive mechanism for quintet formation is yet to be established, which makes it difficult to design materials for optimal quintet formation. Here we outline a mechanism in which inter-triplet exchange coupling fluctuations drive fast and efficient quintet formation. In contrast with conventional wisdom, we show that quintet population can arise despite strong exchange coupling. We evaluate the performance of this quintet formation mechanism in two regimes of conformational freedom, and relate quintet dynamics to material properties of singlet fission molecules.
\end{abstract}

\maketitle
\makeatletter

\section{Introduction}
\label{s:introduction}
\noindent
Exchange interactions between electrons with overlapping wavefunctions arise from exchange symmetry.
Engineered exchange interactions underpin many new technologies including spintronics, quantum information, magnetic materials and spin dependent chemical processes. In many of these applications, precise control of the exchange interaction is also critical. However, as with all realistic systems, these exchange interactions are subject to noise. 

Exchange noise can arise from many sources, including thermally-driven structural fluctuations, switching in nearby charge centres~\cite{Keith2022, Eng2015}, and electrical noise in voltages applied to control gates~\cite{Testolin2009} in engineered systems. Quantum dot systems~\cite{Li2010, Saraiva2022} are a prime example of this and, due to their potential applications in quantum information processing, advanced approaches to quantifying the impact of exchange noise have been developed~\cite{Ferrie2018}. Fundamentally, these approaches have sought to efficiently model open quantum systems (OQS), i.e., quantum systems coupled to their environment~\cite{Breuer2016}, as a way to understand and improve fidelity of gate operations~\cite{Khodjasteh2010, Soare2014}. Interestingly, there are systems where exchange noise, instead of being a detriment, plays a necessary part in useful quantum processes~\cite{Modi2011,Campaioli2018}. In this work, we apply OQS modelling techniques to understand how fluctuation of exchange coupling produces high-spin states in singlet fission (SF).

SF is a photophysical process that occurs in molecular systems, wherein an optically prepared singlet (spin-0) exciton forms a pair of triplet (spin-1) excitons on neighbouring chromophores~\cite{Berkelbach2013}. The process has been the subject of fundamental spectroscopic studies since the 1960s~\cite{Johnson1967,Swenberg1968}, and has received renewed interest this century due to its potential use in photovoltaic devices~\cite{Smith2010,Congreve2013,Einzinger2019,Baldacchino2022} and medical imaging~\cite{kawashima2022}. Since SF proceeds rapidly from a singlet state, it is assumed to form the \textit{net-singlet} triplet-pair $\singlet$, i.e., a pair of triplets whose spins couple with zero net spin. However, recent spectroscopic studies have revealed that the triplet-pair undergoes spin dynamics, forming triplet $\triplet$ and quintet $\quintet$ multiexcitons~\cite{Tayebjee2016,Chen2019}, before dissociating into uncorrelated triplet excitons. High-spin states such as quintets have fundamental implications for the use of SF in photovoltaics~\cite{Chen2019}, and have also been considered for quantum information processing applications~\cite{Jacobberger2022}; understanding how these high-spin states form in SF, and how material design can affect their formation, presents an important unanswered question in the field~\cite{Jacobberger2022}.

One proposed mechanism for high-spin state generation is fluctuation of the inter-triplet exchange coupling~\cite{Collins2019a,Weiss2017}. Given that the exchange coupling strength is sensitive to inter-triplet wavefunction overlap, such fluctuations likely arise from the nuclear motions of the molecules hosting the excitons~\cite{kobori2020}. In a recent work~\cite{Collins2019a}, the authors show that transitions from weak to strong exchange, mediated by conformational motion, offer a pathway for quintet formation. However, such a mechanism requires the exchange coupling to be weak for a sufficiently long time for spin-mixing to occur, on the order of nanoseconds.
\begin{figure*}[ht]
    \centering
    \includegraphics[width=0.98\textwidth]{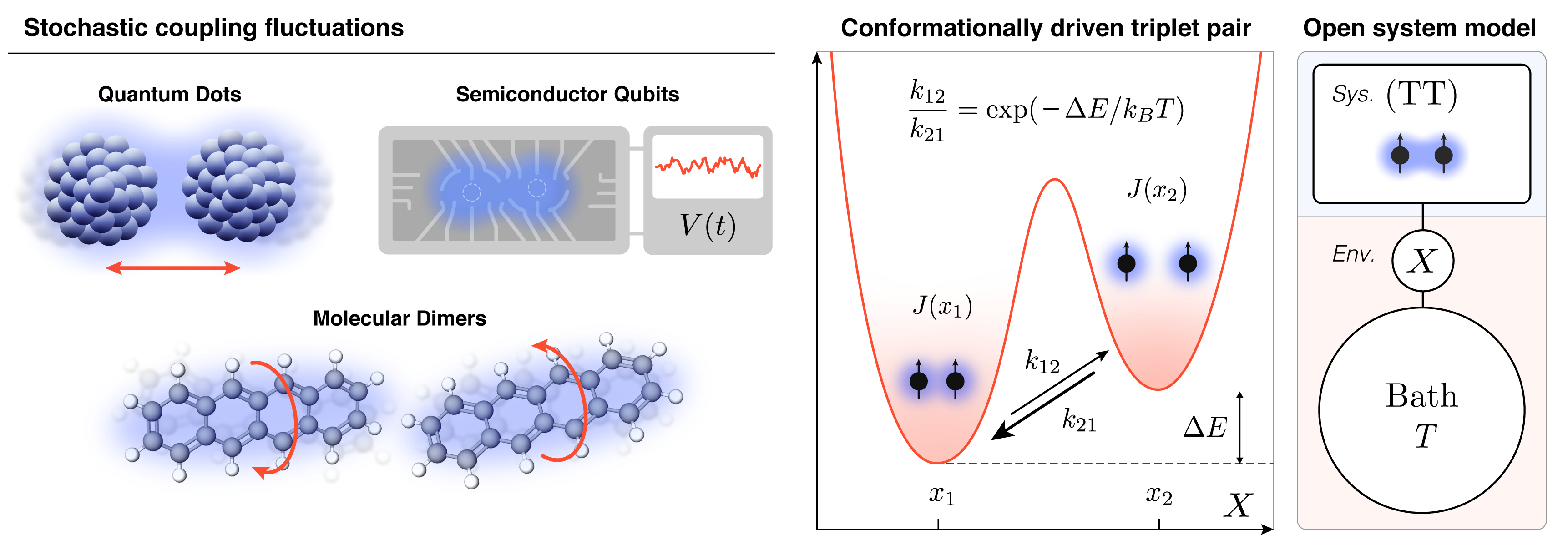}
    \caption{(\textit{Color online}) (\textit{Left}) Exchange coupling fluctuations that arise from various source of noise are ubiquitous. They often negatively affect the performance of quantum information processing of quantum dots and superconducing qubits~\cite{Laucht2021}. In molecular dimers, exchange coupling fluctuations can instead be beneficial, as we show in this work for the formation of high-spin states.
    (\textit{Right}) Here, we consider a system given by a triplet pair $(\TT)$, consisting of two {spin-1} particles, that undergoes spin mixing due to fluctuations in the exchange interaction $J(x_t)$ driven by a conformational coordinate $X$. The latter interacts with a large bath of nuclear vibrations at thermal equilibrium with temperature $T$. The efficiency of quintet formation depends on the stochastic process $\{X(t):t\in\mathbb{R}_+\} \mapsto\{J(x_t)\}$, whose two-time correlation functions respect the thermodynamic detailed balance condition.}
    \label{fig:1_schematic}
\end{figure*}

Meanwhile, experimental evidence suggests that quintet formation may be driven by conformational dynamics even if nuclear reorganisation proceed within picosecond timescales~\cite{kobori2020}. Furthermore, the broadness of EPR spectra, directly linked to the strength of the exchange interaction, provides an additional indication that quintets might form even in the strong-exchange regime~\cite{Tayebjee2016}. Indeed, no magnetic resonance studies of covalent SF dimers to date demonstrate the formation of weakly-coupled high-spin states prior to strongly-coupled high-spin states. All these observations urge clarification and lead us to two essential questions: Can quintet multiexciton formation proceed efficiently even in the strong-exchange regime? And if so, what are the ideal conformational properties---e.g., \textit{looseness} or \textit{stiffness} of the host molecule---for enhancing or inhibiting spin-mixing?

In this work, we systematically address these questions using an open quantum system approach to model the dynamics of the correlated triplet pair $(\TT)$ as it interacts with its environment. By considering the paradigmatic cases of stochastic conformational switching---inspired by the recent experimental work of Ref.~\cite{kobori2020}---and harmonic conformational motion, we show that quintet formation can proceed efficiently even if the exchange interaction is orders of magnitude larger than the coupling between singlets and quintets.
With exact numerical solutions and fundamental results from the theory of open quantum systems we precisely interpret the mechanisms for which conformational dynamics assists high-spin formation. We also present closed-form expressions for the optimal conditions for quintet formation, its dependence on temperature, magnetic field, and noise power spectrum of the conformational dynamics. We conclude by discussing the significance of our results from both fundamental and practical standpoints.

\section{Methodology}
\label{s:methodology}
\noindent
The system considered in this work is the correlated triplet pair $(\TT)$, modelled using the spin Hamiltonian
\begin{equation}
    \label{eq:spin-hamiltonian}
    H_{\TT} = H_\mathrm{z} + H_\mathrm{zfs} + H_\mathrm{ee},
\end{equation}
given by the sum of Zeeman ($\mathrm{z}$), zero-field splitting ($\mathrm{zfs}$) and exchange ($\mathrm{ee}$) interactions~\cite{Goldfarb2018}, as done in Refs.~\cite{Tayebjee2016,Collins2019a,kobori2020}. We ignore the effects of triplet diffusion, by assuming the pair to sit on two neighbouring sites of a dilute crystal or on a molecular dimer~\cite{Nakano2016,Tayebjee2016,Nakano2019,Smyser2020,kobori2020}.
To focus on the strong-exchange regime we set the exchange strength to be much larger than the zero-field splitting and Zeeman interactions, i.e., $\| H_\mathrm{zfs} \| /\|H_\mathrm{ee}\|,\| H_\mathrm{z} \| /\|H_\mathrm{ee}\| \ll 1$, by account of the spectral norm {$\|\cdot\|$}.
Explicit expressions for the terms in the Hamiltonian of Eq.~\eqref{eq:spin-hamiltonian} are given in Sec.~\ref{sm:spin_hamitlonian} of  the Supplemental Material (SM)~\cite{SM}. 

The singlet, triplet, and quintet states (denoted $\singlet$, $\triplet$, and $\quintet$ respectively) are defined as eigenstates of the triplet-pair $S^2$ operator~\cite{SM}. Since both $H_\mathrm{z}$ and $H_\mathrm{ee}$ commute with the total spin operator $S^2$, they cannot mix $\singlet$ with the high-spin states, while $H_\mathrm{zfs}$ can. In Sec.~\ref{s:results} we fix the parameters of $H_\mathrm{zfs}$~\cite{SM} such that triplet excitons are indistinguishable, to prevent the $\singlet$ (symmetric under permutation of triplets) from mixing with $\triplet$ (antisymmetric). As we will discuss in Sec.~\ref{ss:nofield}, this choice remarkably simplifies the rationalisation of the spin dynamics, which can often be reduced to that of a two-level system. Nevertheless, our approach is of general validity and can be applied to arbitrary choices of zero-field splitting parameters~\cite{SM}. 

To study the role of conformational motion on multiexciton dynamics we consider the simplified scenario in which a single conformational coordinate $X$ is responsible for the fluctuations of the exchange interaction strength $J(x_t)$~\cite{Goldfarb2018}, with $x_t$ being the value of $X$ at time $t$. Note that the coordinate $X$ is not necessarily a proxy for physical distance between two sites, and could, for example, represent the asymmetry parameter of a double quantum well or the relative angle between two molecules. A large ensemble of nuclear vibrations, here modelled as a phonon bath at thermal equilibrium, is directly coupled only to $X$ (as shown in the schematic of Fig.~\ref{fig:1_schematic}), driving transitions between different conformational configurations.

Throughout this work, we assume that the dynamics of the conformational coordinate $X$ and that of the bath are not affected by that of the triplet pair, as often done in the literature~\cite{Berkelbach2013b,kobori2020}.
This allows us to study the dynamics of $(\TT)$ in two regimes of conformational dynamics: \textit{Stochastic switching} and \textit{perturbed harmonic oscillations}.

First, we consider the case in which the vibrational bath drives stochastic switching of $X$ between two configurations $x_1$ and $x_2$, as depicted in Fig.~\ref{fig:1_schematic}. This is akin to the systems considered by Kobori \textit{et al.} in Ref.~\cite{kobori2020} and by Korovina \textit{et al.} in Ref.~\cite{Korovina2020}. This model is physically well motivated for stable and thermally accessible configurations $x_1$, $x_2$ (e.g., asymmetric double well potentials), energetically separated by $\Delta E:=E(x_2)-E(x_1) > 0$ (without loss of generality) such that the thermal energy $k_B T$ at temperature $T$ is sufficiently large to induce hopping between the local equilibria~\cite{Gammaitoni1998}. 
In Sec.~\ref{ss:nofield} we use this model to study quintet formation in the strong-exchange regime at zero-field, i.e., in the absence of Zeeman interaction. The effect of magnetic field intensity and orientation is then discussed in Sec.~\ref{ss:field}.

We then consider a continuous conformational space in Sec.~\ref{ss:harmonic_conformation}, where we model $X$ as an harmonic mode with characteristic frequency $\omega$. The mode exchanges energy with the thermal bath at some rate $\gamma(\Delta E,T)$ that respects the thermodynamic detailed balance condition. With this model we aim to study high-spin state formation driven by a conformational coordinate that oscillates around a unique thermally-accessible local equilibrium. By studying the dynamics of $(\TT)$ over the parameter space spanned by $\omega$ and $\gamma_0 = \gamma(0,T)$, we highlight the relation between quintet formation efficiency and the \textit{noise power spectrum} (i.e., noise colour) of the conformational stochastic process $\{X(t):t\in\mathbb{R}_+\}$~\cite{Niepce2021}. The role of the noise memory kernel is then framed in terms of \textit{Markovian} and \textit{non-Markovian}~\cite{Breuer2002,Milz2021a} conformational driving of the spin manifold.

\section{Results}
\label{s:results}

\subsection{Stochastic conformational switching at zero-field}
\label{ss:nofield}

\noindent
Let us consider the system of Eq.~\eqref{eq:spin-hamiltonian} in the absence of an external magnetic field $\bm{B}$ ($H_\mathrm{z} = 0$). The stochastic switching of $X$ affects the strength $J(x_t)$ of the exchange interaction, which takes the value $J_i$ at configuration $x_i$. 
The conformational dynamics is fully described by the rates $k_{ij}$ of switching from configuration $x_i$ to $x_j$. Note that the rates $k_{ij}$ do not depend on the $(\TT)$ states, because the spin dynamics is assumed to not affect that of $X$.

To study the dynamics of the correlated triplet pair we consider the Hilbert space $\mathcal{H} = \mathcal{H}_{\TT,x_1}\oplus \mathcal{H}_{\TT,x_2}$ associated with the system $(\TT)$ at configurations $x_1$ and $x_2$, and rearrange it as $\mathcal{H} = \mathcal{H}_{X}\otimes\mathcal{H}_{\TT}$. The dynamics of the state $\rho \in \mathcal{S}(\mathcal{H})$ is then determined by the following Lindblad master equation
\begin{equation}
\label{eq:master_equation}
    \dot{\rho}_t = \frac{i}{\hbar}[\rho_t,H] + \sum_{\substack{i=1,2\\j\neq i}}k_{ij}\Big(L^{\phantomdagger}_{ij}\rho_t L_{ij}^{\dagger} - \frac{1}{2}\big\{L^{\dagger}_{ij}L_{ij}^{\phantomdagger},\rho_t\big\}\Big),
\end{equation}
where ${\dagger}$ denotes the Hermitian conjugate and $\{\cdot,\cdot\}$ is the anticommutator.
Here, $H = \small\sum_{i=1,2}\Pi_{x_i}\otimes H_{\TT}(x_i)$, where $\Pi_{x_i} = \ketbra{x_i}{x_i}$ is the projector on configuration $x_i$, and $H_{\TT}(x_i)$ is the Hamiltonian of Eq.~\eqref{eq:spin-hamiltonian} evaluated at configuration $x_i$. Similarly, the Lindblad (\textit{jump}) operators $L_{ij} = \ketbra{x_j}{x_i}\otimes\mathbb{1}_{\TT}$ model stochastic conformational switching $x_i\to x_j$ at rate $k_{ij}$, without acting on the state of the correlated triplet pair. Eq.~\eqref{eq:master_equation} provides a Markovian description of the $\TT$ dynamics averaged over the ensemble of all possible conformational trajectories. Note that this approach is non-perturbative in $H_{\TT}(x_i)$, so that we can consider arbitrary dependence of the spin Hamiltonian on $X$.

Eq.~\eqref{eq:master_equation} is solved using the Liouville superoperator approach $\dot{\bm{\rho}}_t = \mathcal{L}\bm{\rho}_t$ to obtain $\bm{\rho}_t = \exp[\mathcal{L} t]\bm{\rho}_{0}$, where $\mathcal{L}$ is the Liouville superoperator associated with Eq.~\eqref{eq:master_equation}, and where the initial state $\rho_{0} = \ketbra{x_2}{x_2}\otimes\ketbra{\singlet[0]}{\singlet[0]}$ is assumed to be the singlet state $\singlet[0]$ at the high-energy configuration $x_2$. This choice reflects the intention of studying SF as a non-equilibrium process that, following photoexcitation, proceeds via thermal relaxation starting from an out-of-equilibrium state of $(\TT)$ and $X$~\cite{Berkelbach2013b}. 

\begin{figure}[t]
    \centering
    \includegraphics[width=0.48\textwidth]{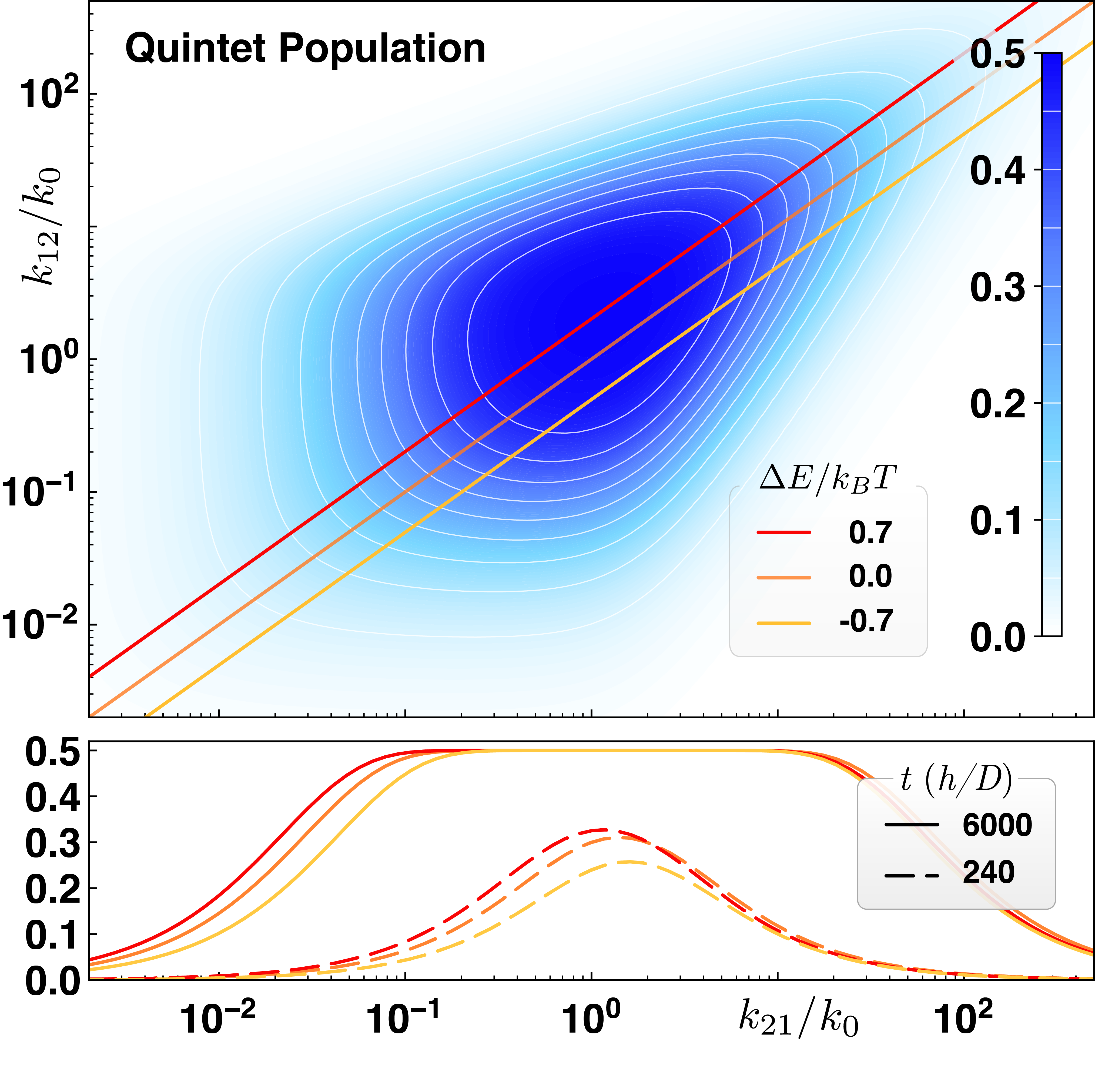}
    \caption{(\textit{Colour online}) \textbf{Quintet formation driven by stochastic conformational switching at zero-field.}---(\textit{Top}) Quintet population $p_5(t)$ for $t =1200$ periods ($h/D$,) or $\approx1.05\mu$s, as a function of the switching rates $k_{12}$, $k_{21}$ in units of $k_0 = 3(J_1+J_2)/2h$, where $h$ is the Planck constant. Quintet formation proceeds efficiently when the stochastic resonance conditions of Eq.~\eqref{eq:optimal_stochastic_switching} is respected. (\textit{Bottom}) Slices of $p_5(t)$ at $k_{21}/k_{12} = \exp[\Delta E/k_B T]$ for $\Delta E/k_B T = -0.7$, $0$, and $0.7$, for $t = 240,6000$ periods ($h/D$) as indicated in the legend. The stochastic resonance condition for rates is not generally symmetric in $k_{ij}$.
    Stochastic resonance drives quintet formation by switching between two non-commuting spin Hamiltonians. At the individual trajectory level $\singlet$ and $\quintet$ can undergo complete population inversion. However, the ensemble average of $p_5(t)$ never exceeds $1/2$. This is because for sufficiently long times  each individual spin trajectory ergodically explores the space of accessible states driving the ensemble towards the state of maximal entropy. See Sec.~\ref{sm:spin_hamitlonian} for parameters of the spin Hamiltonian~\cite{SM}.}
    \label{fig:2_nofield}
\end{figure}

Our results, presented in Fig.~\ref{fig:2_nofield} for a particular choice of zero-field splitting and exchange parameters~\cite{SM}, show the average population $p_5(t)$ of the quintet manifold $\quintet[m]$
\begin{equation}
    \label{eq:quintet_population}
    p_5(t) = \sum_{m = -2}^{2}\tr[\rho_t\:\mathbb{1}_{X}\otimes \ketbra{\quintet[m]}{\quintet[m]}],
\end{equation}
as a function of the switching rates $k_{ij}$. From our solution, it is evident that quintet formation can proceed \textit{efficiently}---i.e., with significant quintet/singlet population ratios---and \textit{rapidly}---i.e., within the $\mathrm{ns}$ to $\mathrm{\mu s}$ timescale characteristic of high-spin lifetimes in EPR experiments~\cite{Pun2019}---even in the strong-exchange regime. More importantly, our results provide a prescription for the optimisation of the conformational switching parameters $k_{ij}$ for enhancing quintet formation.

The mechanism of spin-mixing presented in Fig.~\ref{fig:2_nofield} can be rationalised by representing the correlated triplet pair in terms of an equivalent two-level system (TLS). In the absence of a magnetic field, and assuming parallel chromophores~\cite{SM}, the initial singlet state $\ket{\singlet}$ only couples with a unique accessible state with quintet character, here $\ket{\quintet}$ for brevity. Under these conditions the spin Hamiltonian can be rewritten as
\begin{equation}
    \label{eq:equivalent_TLS}
    H_\mathrm{TLS} = - \frac{\Delta}{2}\sigma_x - \frac{\varepsilon(x_t)}{2}\sigma_z,
\end{equation}
where $\sigma_x$ and $\sigma_z$ are Pauli operators. Here  $\Delta = -2\braket{\singlet|H_{\TT}|\quintet}$ is directly associated with the zero-field splitting parameters coupling $\ket{\singlet}$ and $\ket{\quintet}$, while $\varepsilon(x_t) = \braket{\quintet|H_\TT|\quintet}- \braket{\singlet|H_\TT|\singlet}$ is related to the strength $J(x_t)$ of the exchange interaction and the zero-field splitting parameters; see Sec.~\ref{sm:TLS} of the SM for the explicit expressions of $\Delta$ and $\varepsilon$~\cite{SM}. 

The spin dynamics can now be represented on the Bloch sphere~\cite{Bengtsson2006}: An initial singlet state $\ket{\singlet} \leftrightarrow -\hat{\bm{z}}$ precesses around the axes $\bm{h}_i = \Delta \hat{\bm{x}}+\varepsilon(x_i)\hat{\bm{z}}$ at frequency $\omega_i = \|\bm{h}_i\|/\hbar = \sqrt{\Delta^2+\varepsilon(x_i)^2}/\hbar$. 
In the strong-exchange regime $\Delta \ll \varepsilon(x_t)$, the quickest way to reach the quintet state $\ket{\quintet} \leftrightarrow\hat{\bm{z}}$ is to switch between conformations $x_i\to x_j$ after time intervals $\tau_{ij} \simeq \pi/\omega_i$, i.e., in \textit{resonance} with each conformation's precession frequency. We direct the reader to Sec.~\ref{sm:bloch-sphere-argument} of the SM for a detailed discussion of this geometric argument. 

In the case of stochastic conformational switching (here, with exponential distribution), quintet formation is enhanced when the switching rates $k_{ij}$ match the precession frequencies, so that the average switching time $\braket{\tau_{ij}} \approx \pi/\omega_{i}$. This relation, well-known as the statistical synchronisation condition~\footnote{Stochastic resonance occurs when the average switching time matches half the period of periodic driving~\cite{Gammaitoni1998}.}, allows us to pinpoint stochastic resonance~\cite{Lofstedt1994,Gammaitoni1998} as the fundamental mechanism responsible for efficient quintet formation in the strong-exchange regime. The ideal conditions for enhancing quintet formation are therefore to be sought using
\begin{equation}
    \label{eq:optimal_stochastic_switching}
    k_{ij} = \frac{\sqrt{9 J_i^2-4J_i D + 4 D^2}}{\hbar\pi},
\end{equation}
where $\Delta$ and $\varepsilon$ have been expressed in terms of $J_i = J(x_i)$ and $D$, the relevant exchange and zero-field splitting parameters (See Eqs.~\eqref{eq:Delta},~\eqref{eq:epsilon} in the SM~\cite{SM}). In the limit of strong exchange $J(x_t)\gg D$ the condition reduces to $k_{ij} = 3 J_i/\hbar\pi$.

This observation has two major implications: First, it points at the opportunity of exploiting conformational dynamics with peaked stochastic switching distribution (e.g., Poisson statistics) to further refine the resonance condition, even in the Markovian limit. As we discuss in Sec~\ref{ss:harmonic_conformation}, this leads to further enhancements in quintet formation rate. 

Second, it opens the doors towards the coherent preparation of high-spin states, by means of controlled switching of the exchange interaction strength.
This may be achieved by applying an electric field to modulate the overlap of the electronic wave functions~\cite{Li2008}, or by means of conformational switching. Optically controlled conformational switching, or photoswitching, is a pioneering approach to control the chemical and optical properties of molecular materials~\cite{Helmy2014,Ito2016,Macdonald2021}. Since photoswitching is in itself a stochastic process, the feasibility of photoswitching-assisted high-spin preparation depends on our ability to match the switching rates to the spin-mixing resonance conditions, requiring further refinement of dynamical modelling. In Sec.~\ref{sm:bloch-sphere-argument} of the SM we present an elementary protocol for a switching sequence that can be used to prepare a specific quintet state. We anticipate this approach to receive significant attention for its potential applications.

Before looking at the influence of magnetic field on the spin-mixing dynamics, let us discuss how our results can be used to guide the design of SF materials~\cite{Kumarasamy2017}. Imposing the detailed balance condition on the conformational switching rates $k_{ij}/k_{ji} = \exp{[-\Delta E / k_B T]}$ provides a proxy for the temperature-dependence of the quintet formation rate. We can now interpret our results in terms of \textit{stiffness} and \textit{looseness} of the conformational coordinate: When the energetic separation $\Delta E$ between the stable configurations $x_1$ and $x_2$ is much larger than the thermal energy of the bath (i.e., $X$ is \textit{stiff}) quintet formation is inhibited. The same outcome is expected for sufficiently low temperatures $T$, such that the coordinate $X$ is considered to be \emph{frozen}. This is consistent with the experimental findings of Ref.~\cite{kobori2020}, where a fast-oscillating conformational switching ($\mathrm{THz}$) is identified as responsible, over other slower modes, for quintet formation following SF in TIPS-pentacene molecular dimers. Note that spin-mixing is inhibited also when both rates are too slow, too fast, or generally away from the stochastic resonance condition. 

\subsection{Magnetic field effects on stochastic conformational switching}
\label{ss:field}
\noindent 
Magnetic field experiments are crucial for identifying the effects of spin on organic electronic processes such as SF. The assumption of no field in Sec 3A simplifies the calculation, but prevents comparison between this theory and the various spin-probing experiments such as EPR and ODMR. In this section we explore how a static magnetic field $\bm{B}$, with variable magnitude $B=|\bm{B}|$ and orientation relative to the molecule $\hat{e}_B = \bm{B}/B$, affects quintet formation in the stochastic switching model of Sec \ref{ss:nofield}.

Note that the field parameters here are simplified by two assumptions: our dimers are coplanar, and so share a common z-axis $\hat{z}_M$; and the ZFS parameter $E$ is set to zero~\cite{SM}. Hence our molecule spin Hamiltonian has $D_{\infty h}$ symmetry (i.e it is invariant under rotations about $\hat{z}_M$,) so the field direction $\hat{e}_B$ is uniquely determined by the angle $\theta$ between $\hat{e}_B$ and $\hat{z}_M$.

\begin{figure}
    \centering
    \includegraphics[width=0.46\textwidth]{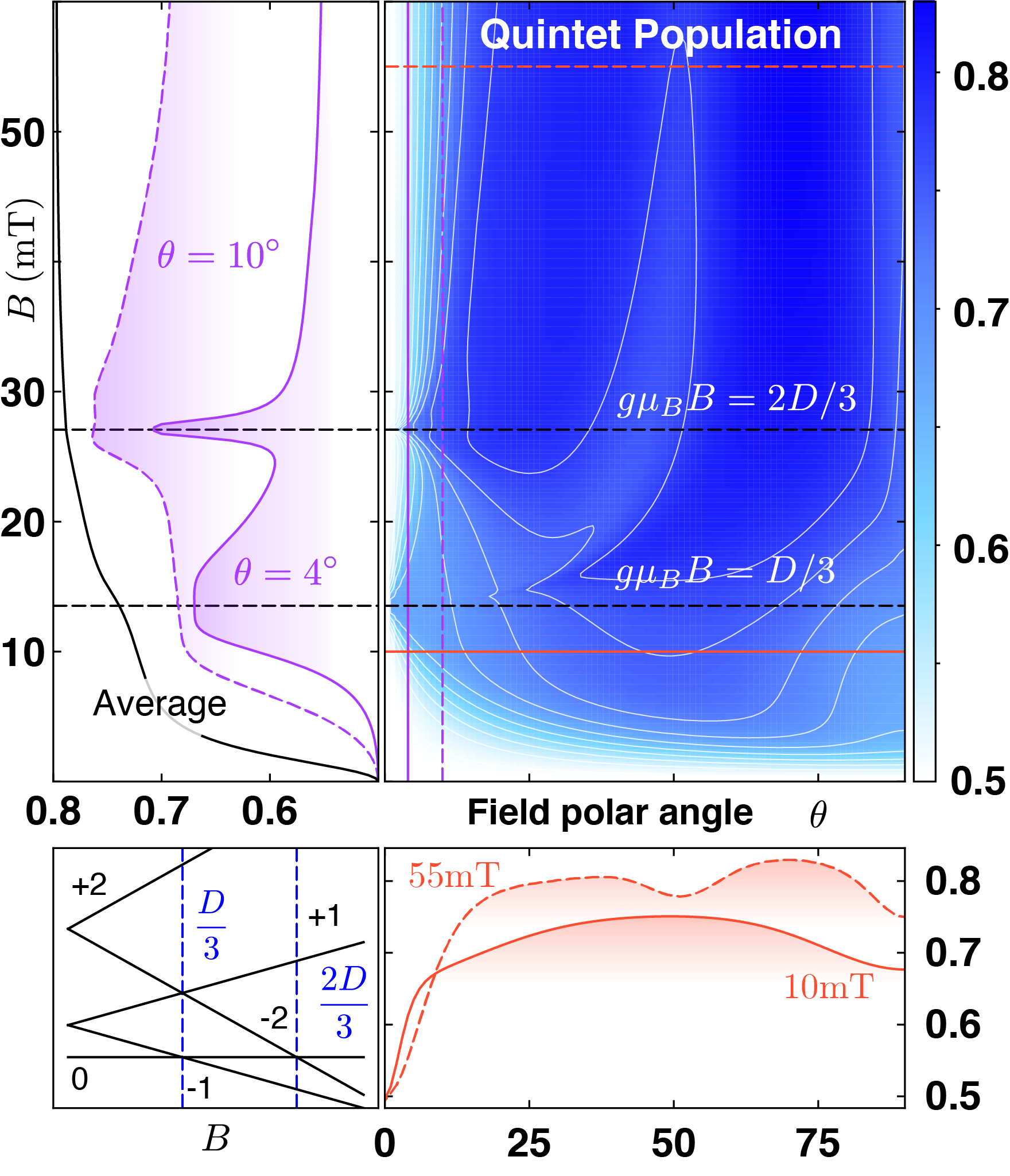}
    \caption{(\textit{Colour online}) \textbf{Effect of magnetic field on quintet formation driven by stochastic conformational switching.}---(\textit{Top-right}) Quintet population $p_5(t)$ at $t = 2000h/D\approx1.76\:\mathrm{\mu s}$ as a function of the field polar angle $\theta$ and the magnetic field strength $B$. (\textit{top-left}) Slices of $p_5(t)$ for fixed angles as a function of $B$, with the weighted average of $p_5$ over all angles in black. The peaks in the $\theta=4^{\circ}$ slice correspond to the level crossings shown in the bottom left. (\textit{Bottom-right}) Slices of $p_5(t)$ for fixed field strength as a function of $\theta$. The dips in the at $\theta=0^{\circ}$, $54.7^{\circ}$, and $90^{\circ}$ occur where some $\quintet$ eigenstates cannot mix with $\singlet$~\cite{Smyser2020}. (\textit{Bottom-left}): Quintet energy levels varying with magnetic field strength, with field direction $\hat{\bm{e}}_B$ parallel to the molecular $z$-axis $\hat{z}_M$. Any deviation in $\hat{\bm{e}}_B$ causes the crossings to become avoided, allowing more than one $\quintet$ state to mix with $\singlet$. See Sec.~\ref{sm:spin_hamitlonian} for parameters of the spin Hamiltonian~\cite{SM}.}
    \label{fig:3_field}
\end{figure}

Changing the field orientation changes the symmetry of the spin Hamiltonian $H_{\textrm{TT}}$, which affects the number of $\quintet[]$ states that are accessible from $\singlet$~\cite{Smyser2020}. In Fig.~\ref{fig:2_nofield}, when only one $\quintet[]$ state can be accessed from $\singlet[]$, the $\quintet[]$ population saturates at 50\%. If $\singlet$ were able to mix with more of the $\quintet$ states, two things should occur: The $\quintet$ population should saturate at higher percentages; and the net rate of $\quintet$ population should increase, as $\quintet$ formation becomes entropically favourable. These magnetic field effects are qualitatively similar to those studied by Merrifield et al.~\cite{Merrifield1968} to explain the effects of magnetic field on triplet excitons dynamics.

Fig.~\ref{fig:3_field} shows the quintet population at $t=2000\:h/D\approx1.76\mu$s, which is a sufficiently long time for the quintet population to nearly saturate for every value of the field parameters. Fig.~\ref{fig:3_field} (\textit{top-right}) shows how the quintet population varies with $\theta$ and $B$. Both extremes of $H_{\textrm{TT}}$ symmetry can be seen: there are regions where only one $\quintet$ state mixes with $\singlet$, so the quintet population saturates at 0.5; and regions where all five $\quintet$ states mix, and quintet population saturates at $5/6\approx0.83$. While the two-conformation stochastic model is simple, it allows us to draw qualitative conclusions about how quintet formation rate varies with field direction. This information may be fruitfully coupled with EPR and ODMR experiments, which are able to detect quintet population in a way that is sensitive to field direction~\cite{Yunusova2020}. For example, at high field the region of the EPR/ODMR spectrum corresponding to $\theta=50^{\circ}$ is predicted here to form quintets more slowly than, say, $\theta=70^{\circ}$.

Fig.~\ref{fig:3_field} (\textit{top-left}) shows quintet population as a function of $B$ for $\theta=4^{\circ}$, $\theta=10^{\circ}$, and averaged over all $\theta$ values. There are two pronounced peaks when $\theta$ is small, coinciding with level crossings of the $\quintet$ sublevels as shown in Fig.~\ref{fig:3_field} (\textit{bottom-right}). These level crossings are plotted for $\theta=0$, in which case the crossings are not avoided, but small variations in $\theta$ cause the crossings to become avoided while only slightly changing where the crossings occur. While these crossings occur far below conventional EPR fields, they may be visible in magneto-photoluminescence experiments, which can indirectly observe $\quintet$ formation as a lack of $\singlet$ absorption signal.

Lastly, Fig.~\ref{fig:3_field} (\textit{bottom-right}) shows quintet population varying with $B$ while $\theta$ is held constant. It is clear that when $\theta=0$, i.e $\bm{B}$ is aligned with the molecular $z$-axis $\hat{z}_M$, only one $\quintet$ state can mix with $\singlet$. $\theta=90^{\circ}$ also shows less than the maximal quintet yield, with $\quintet$ populations saturating at 3/4, since only three $\quintet$ states mix with $\singlet$ as predicted by~\cite{Smyser2020}.

The results of Section \ref{ss:field} show that even if the magnetic field is weak compared to exchange coupling ($\|H_{\textrm{z}}\|/|H_{\textrm{ee}}\|\approx 0.01$,) it can greatly affect the rate of quintet formation in the stochastic switching model. Firstly, this suggests magnetic field as a valuable control parameter for quintet generation, supplementing the conformational properties proposed in Section \ref{ss:nofield}. Secondly, it allows for the results of already common experimental measurements of quintet formation, such as EPR and ODMR spectroscopy, to be interpreted without assuming weak exchange coupling~\cite{Tayebjee2016}.

\subsection{Harmonic conformational dynamics at zero-field}
\label{ss:harmonic_conformation}

\noindent
We now generalise our results to the case of a continuous conformational space, by modelling $X$ as a harmonic mode coupled to a thermal bath. For the sake of clarity, we go back to the spin Hamiltonian considered in Sec.~\ref{ss:nofield}, by setting $H_\mathrm{z} = 0$. This choice simplifies the solution of the spin dynamics and the interpretation of the results, by reducing the triplet-pair to the equivalent TLS of Eq.~\eqref{eq:equivalent_TLS}.

The conformational coordinate $X$, whose dynamics is still independent of the state of the spins, is here modelled as a harmonic oscillator with characteristic frequency $\omega$, that exchanges energy $\Delta E$ with the bath at temperature $T$ at rate $\gamma(\Delta E, T)$ such that $\gamma(0,T) = \gamma_0$, here referred to as the \textit{noise power}. The trajectories $x_t$ of $X$ correspond to harmonic oscillation $A\cos(\omega t + \phi)$ around an equilibrium $x_0 = 0$ (without loss of generality), intermitted by stochastic changes of phase and amplitude $(A,\phi)\to(A',\phi')$; they are sampled numerically as described in Sec.~\ref{sm:sampling_harmonic_conformational_dynamics} the SM~\cite{SM}.

Dealing with a continuous conformational space, we now study the spin dynamics using the time-dependent Schr\"odinger equation $i\hbar d_t\ket{\psi_t} = H_\TT(t)\ket{\psi_t}$, where the Hamiltonian $H_\TT(t)$ depends on conformational trajectories via $J(x_t)$---here, a linear function of $x_t$---and where $\ket{\psi_t}$ is the state of the correlated triplet pair time $t$. The solution $\ket{\psi_t} = U(t,t_0)\ket{\psi_0}$ along a trajectory $x_t$ can be expressed in terms of the Dyson series $U(t,t_0) = \mathcal{T}\{\exp[-i\int_{t_0}^t ds H_\TT(s)]\}$~\cite{Breuer2002}. The spin dynamics $\rho_t$ of the ensemble is then obtained by averaging over a large number of trajectories, as discussed in the SM~\cite{SM}.
\begin{figure}
    \centering
    \includegraphics[width=0.46\textwidth]{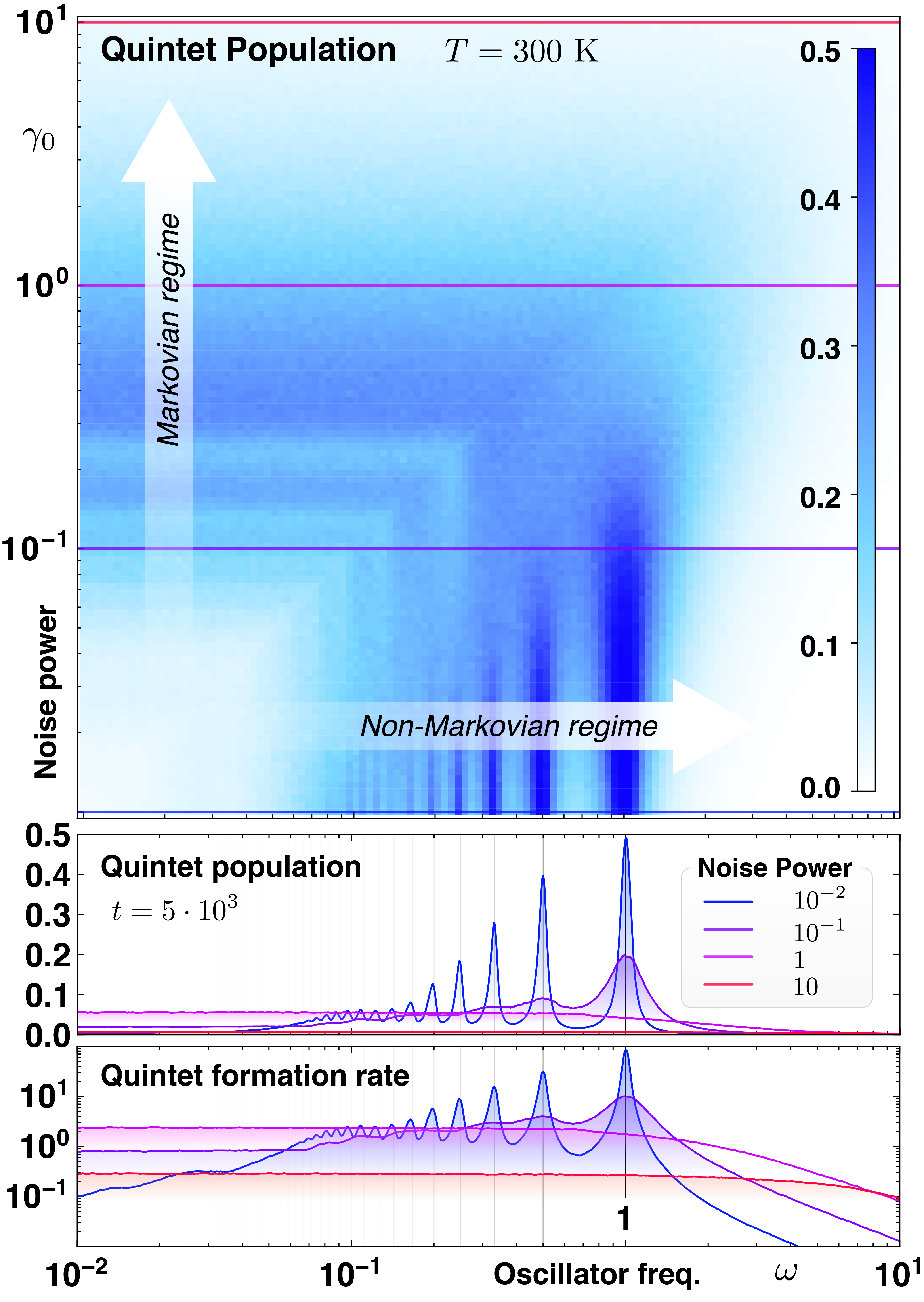}
    \caption{(\textit{Color online}) \textbf{Quintet formation driven by harmonic conformational dynamics.}---(\textit{Top}) Quintet population for $t = 5000$ periods ($\hbar/\varepsilon_0$), at $T = 300~\mathrm{K}$, as a function of the oscillator frequency $\omega$ and noise power $\gamma_0$ (frequencies and rates expressed in units of $\varepsilon_0/\hbar$, 1000 trajectories for each point).  Quintet population (\textit{middle}) and quintet formation rate (\textit{bottom}) along slices of constant noise power $\gamma_0$ (see legend); rates $\Gamma$ from model $p_5(t) = [1-\exp(-\Gamma t)]/2$.  Pronounced resonance peaks can be seen for $\gamma_0\ll\omega$ (\textit{non-Markovian regime}) at $\omega = \varepsilon_0/k$, with $k$ positive natural numbers. Their positions is prescribed by the diabatic limit ($\omega\gg\Delta^2/A$) of multiple-passage Landau-Zener-St\"uckelberg theory~\cite{Shevchenko2010}. Increasing the noise power $\gamma_0$ decreases the coherence lifetime of the triplet pair, so there is less opportunity for the St\"uckelberg phase of the triplet pair to accumulate over several periods of oscillation. This results in the resonance peaks broadening to the point of indistinguishability for moderate noise power $\gamma_0\gg\omega$ (\textit{Markovian regime}). See Sec.~\ref{sm:spin_hamitlonian} for parameters of the spin Hamiltonian~\cite{SM}.}
    \label{fig:4_harmonic}
\end{figure}

As done in Sec.~\ref{ss:nofield}, we focus on the average population of the quintet manifold $p_5(t)$, here calculated by dropping the $\mathbb{1}_X$ term from Eq.~\eqref{eq:quintet_population}. In Fig.~\ref{fig:4_harmonic} we show how $p_5(t)$ depends on the conformational frequency $\omega$ and on the noise power $\gamma_0$, by fixing the temperature of the bath $T$. As anticipated in Sec.~\ref{ss:nofield}, relaxing the condition of purely stochastic conformational dynamics ($x_{i}\to x_{j}$ at rate $k_{ij}$), we allow for non-trivial correlation timescales for the conformational trajectories, $\braket{x_{t'}x_{t}}\;
\cancel{\propto}\;\delta(t'-t)$~\footnote{White noise is defined as a stochastic process with zero mean and trivial two-time correlation functions $\braket{x_{t'}x_{t}} \propto \delta(t'-t)$.}, thus opening up to the possibility of partially coherent resonant driving of the correlated triplet pair. These leads to enhanced resonance conditions that can outperform the quintet formation efficiencies typical of stochastic resonance.

This fundamental difference from the results of Sec.~\ref{ss:nofield} is highlighted by the formation of the \textit{resonance} regions for low noise powers $\gamma_0\ll\omega$ shown in Fig.~\ref{fig:4_harmonic}. The peak of the resonance fringes can be found at $\omega_k = \varepsilon(x_0)/k$ (c.f Eq.~\eqref{eq:equivalent_TLS},) for positive natural numbers $k$, as prescribed by the diabatic limit of the Landau-Zener-St\"uckelberg theory of periodically driven~\footnote{Here, acoustically driven.} two-level systems~\cite{Shevchenko2010}. Larger resonance frequencies $\omega_k$ induce faster mixing, and therefore higher peaks, due to the higher number of LZS \textit{passages}~\cite{Shevchenko2010}. Note that the LZS theory, based on the Hamiltonian of Eq.~\eqref{eq:equivalent_TLS} for the case of harmonic oscillations of $\varepsilon(t) = \varepsilon_0 + A \sin(\omega t)$, provides closed form solutions only for the resonance conditions in the slow-driving (\textit{adiabatic}) and fast-driving (\textit{diabatic}) regimes, given by $A\omega\ll\Delta^2$ and $A\omega\gg\Delta^2$, respectively. Our exact numerical solutions complement the theory while demonstrating the robustness of the resonance conditions outside the limit cases.

Once again, these results have significant implications for the design of singlet fission materials, as the formation of high-spin states can be tuned by means of resonant (or off-resonant) conformational motion. The detailed balance condition $\gamma(\Delta E,T)/\gamma(-\Delta E,T) = \exp[\Delta E/k_B T]$ can be used to determine the ideal \textit{stiffness} or \textit{looseness} of the conformational environment to achieve the desired spin-mixing, suggesting that resonance fringes would be more evident at low temperatures, for $X$ weakly coupled to the thermal bath.

Finally, we would like to remark how the noise power spectrum $S_J(\omega) = \mathcal{F}[c_J(t+\tau,t)]$ of the stochastic process associated with the time variation of exchange strength $J(t)$, given by the Fourier transform of its autocorrelation function $c_J(t+\tau,t) = \braket{J(t+\tau)J(t)}$, provides sufficient information to determine if, and how well, the resonance conditions are met. It further highlights the difference between stochastic resonance, which occurs for purely Markovian processes characterised by trivially correlated white noise, and the stronger, partially coherent resonance that arises for processes with coloured noise power spectrum profiles.

\section{Conclusions}
\label{s:conclusions}
\noindent
In this letter we have shown how high-spin states can be generated efficiently in singlet fission even in the strong-exchange regime, when driven by favourable conformational dynamics. This spin-mixing mechanism, here solved for the paradigmatic cases of stochastic switching and harmonic oscillations, is fundamentally different from the one described in Ref.~\cite{Collins2019a}, where quintet formation proceeds in the nanoscecond timescale via conformational reorganisations that briefly suppress the exchange interaction. In practice, we expect these two mechanisms to coexist, and to be experimentally discernible via the different lifetimes and coherence timescales of the generated quintets.

The stochastic and coherent resonance conditions derived in Sec.~\ref{s:results} are key for the engineering of singlet fission materials. They can be used as guidelines to enhance or inhibit the formation of high-spin states, depending on whether they are beneficial or detrimental for tasks like exciton transport and spin-mediated spectral conversion. Our findings also open the path to spin-selective state preparation by coherently switching the exchange interaction strength, which may be controlled with electric fields~\cite{Li2008}, or via conformational photoswitching~\cite{Macdonald2021}. This can impact quantum information processing architectures such as that of Ref.~\cite{Takui2016,Smyser2020}, and help in the experimental interrogation of the fundamental physics of high-spin states, such as their lifetime and interactions with the bath of nuclear spins~\cite{Bayliss2020}.

The considered open quantum system formulation of singlet fission, similar to that of Refs.~\cite{Berkelbach2013,Berkelbach2013b,Mardazad2021}, is also powerful tool for the quantitative study of spin-mixing in singlet fission for specific materials, and can be generalised to account for spin-orbit coupling, spin migration, and other phenomena. It also provides the natural mathematical framework to implement coherent control tasks like fiducial state preparation, that can be tackled using quantum optimal control protocols~\cite{Morley2012,Goodwin2018,Jeschke2019}, and, possibly, saturate bounds for time-optimal state preparation~\cite{Campaioli2018,Campaioli2019a}. By ignoring the effects of triplet diffusion, our results hold for the case of molecular dimers arrangements and dilute crystalline materials with sufficiently low triplet exciton mobility~\cite{Smyser2020}. More complex material like 2D and 3D spin lattices with higher triplet mobility can instead be efficiently addressed using an analogue formulation based on density matrix renormalisation group and tensor network methods~\cite{Jaschke2018, Montangero2018}.

\begin{acknowledgments}
This research was funded in part by the Australian Research Council under grant number CE170100026. This work was conducted using the National Computational Infrastructure (NCI), which is supported by the
Australian Government. M.I.C acknowledges the support of the Sydney Quantum Academy.
\end{acknowledgments}

\bibliography{library,library_manual}
\newpage
\clearpage
\newpage

\begin{center}
\textbf{\large Supplemental Material of \\``\mytitle''}
\end{center}
\setcounter{equation}{0}
\setcounter{figure}{0}
\setcounter{table}{0}
\setcounter{page}{1}
\renewcommand{\theequation}{S\arabic{equation}}
\renewcommand{\thefigure}{S\arabic{figure}}
\renewcommand{\bibnumfmt}[1]{[S#1]}
\renewcommand{\citenumfont}[1]{#1}

\subsection{Spin Hamiltonian}
\label{sm:spin_hamitlonian}
\noindent
The Hamiltonian $H_\mathrm{z}$ is the sum of two local Zeeman interaction terms for each spin subsystem $\nu = 1,2$
\begin{align}
    \label{eq:zeeman_hamiltonian}
    H_\mathrm{z} &= \mu_B g_0 \sum_{\nu = 1,2} \bm{B}\cdot \bm{S}^{(\nu)}, \\
    & = \mu_B g_0 \sum_{i = x,y,z} B_i \bigg( S^{(1)}_i \otimes \mathbb{1}^{(2)} + \mathbb{1}^{(1)} \otimes  S^{(2)}_i \bigg),
\end{align}
where $\mu_B$ is the Bohr magneton, $g_0$ is the scalar Land\'e g-factor. Note that here we are assuming isotropic diagonal $\bm{g}$ tensor; see Ref.~\cite{Collins2019a} in the main text for the general expression.

The dipolar zero-field splitting Hamiltonian $H_\mathrm{zfs}$ is also a sum of local terms $H_\mathrm{zfs} = H_\mathrm{zfs}^{(1)}\otimes\mathbb{1}^{(2)}+ \mathbb{1}^{(1)}\otimes H_\mathrm{zfs}^{(2)}$. Each terms depend on the $\bm{D}$ tensor
\begin{equation}
    \label{eq:D-tensor}
    \bm{D}^{(\nu)} = R_{\nu} \bm{D}_0 R^{T}_{\nu},
\end{equation}
where $\bm{D}_0 = \mathrm{diag}(-D/3+E,-D/3-E,2D/3)$, with $D,E$ the being the zero-field splitting parameters, for a give rotation matrix 
\begin{equation}
    \label{eq:rotation_matrix}
    R_{\nu} = R(\bm{\varphi}_{\nu}) R(\bm{\varphi}_B),
\end{equation}
such that $R(\bm{\varphi}) = R_z(\gamma) R_y(\beta) R_z(\alpha)$ is in the \textit{zyz} convention of the three Euler angles $\bm{\varphi} = (\gamma,\beta,\alpha)$. Each zero-field splitting term reads
\begin{align}
    \label{eq:zfs_hamiltonian}
    H_\mathrm{zfs}^{(\nu)} &= \sum_{i,j = x,y,z} S_i^{(\nu)} D_{ij}^{(\nu)}  S_j^{(\nu)}.
\end{align}

Finally, the intertriplet exchange interaction $H_\mathrm{ee}$ is given by
\begin{equation}
\label{eq:exchange}
    H_\mathrm{ee} = J(x_t) \sum_{i,j = x,y,z} S_i^{(1)}\otimes S_j^{(2)}.
\end{equation}
Note that we assume the exchange interaction to be isotropic. See Ref.~\cite{Collins2019a} in the main text  for the general expression.

\begin{table}[ht]
\begin{tcolorbox}[tabulars*={\renewcommand\arraystretch{1.2}}%
{@{\extracolsep{\fill}\hspace{5mm}}l@{\hspace{5mm}}|l@{\hspace{5mm}}|l@{\hspace{3mm}}},adjusted title=flush left,halign title=flush left,
boxrule=0.5pt,title = {\textbf{Exciton Hamiltonian parameters}}]
\textit{Parameter} & Fig.~\ref{fig:2_nofield},~\ref{fig:3_field} &  Fig.~\ref{fig:4_harmonic} \\
\hline\hline
$D,E$ & $1380,0~\mathrm{MHz}$ & $1380,0~\mathrm{MHz}$ \\ \hline
$\alpha,\beta,\gamma$  & $0,0,0$ & $0,0,0$ \\ \hline
$J_1$ & $5\cdot10^5~\mathrm{MHz}$ & $10^5~\mathrm{MHz}$   \\ \hline
$J_2$ & $1.4\cdot10^5~\mathrm{MHz}$ & $10^4~\mathrm{MHz}$  \\ \hline
\end{tcolorbox}
    \caption{Exciton Hamiltonian parameters used in Fig.~\ref{fig:2_nofield},~\ref{fig:3_field}, and~\ref{fig:4_harmonic}. Energy units expressed in $MHz$, conversion to electronvolt given by $\mathrm{MHz} = 4.13\cdot10^{-9}~\mathrm{eV}$.}
    \label{tab:hamiltonian_parameters}
\end{table}

\subsection{Equivalent Two-level system}
\label{sm:TLS}
\noindent
At zero field ($\|\bm{B}\| = 0$) and symmetric zero-field splitting parameters ($\bm{\varphi}_{1} = \bm{\varphi}_{2}$), the singlet state $\ket{\singlet}$ only couples with a unique quintet state, here $\ket{\quintet}$, which form a two-level system with Hilbert space spanned by a basis $\mathcal{B}:=\{\ket{0},\ket{1}\}$. In this case the spin Hamiltonian $H_TT$ of Eq.~\eqref{eq:spin-hamiltonian} can be reduced to a TLS Hamiltonian
\begin{align}
    \label{eq:tls_equivalent_H}
    H_\mathrm{TLS} &= \begin{pmatrix}
    -2J & \frac{-2\sqrt{2}}{3}D\\
    \frac{-2\sqrt{2}}{3}D & J - \frac{2}{3}D
    \end{pmatrix}, \\
    &=\bigg(-\frac{3J}{2}+\frac{D}{3}\bigg)\sigma_z - \frac{2\sqrt{2}D}{3}\sigma_x
        - \bigg(\frac{J}{2}+\frac{D}{3}\bigg)\mathbb{1}.
\end{align}
First, we drop the term proportional to $\mathbb{1}$, which does not affect the dynamics of singlet and quintet populations. Then we write the TLS Hamiltonian in terms of two parameters $\Delta$ and $\varepsilon(x_t)$, given by
\begin{align}
    \label{eq:Delta}
    &\Delta = \frac{4\sqrt{2}}{3} D, \\
    \label{eq:epsilon}
    &\varepsilon(x_t) = 3J(x_t) -\frac{2}{3}D,
\end{align}
that only depend on the exchange strength $J(x_t)$ and zero-field splitting parameter $D$.
The resulting Hamiltonian reads
\begin{equation}
    H_\mathrm{TLS}(t) = -\frac{\Delta}{2}\sigma_x - \frac{\varepsilon(x_t)}{2}\sigma_z.
\end{equation}

In Sec.~\ref{ss:harmonic_conformation} we consider the case of harmonic conformational oscillation, leading to the following sinusoidal exchange interaction strength,
\begin{equation}
\label{eq:LZS_sinusoidal}
    J(t) = \varepsilon_0 + A\sin(\omega t),
\end{equation}
where 
\begin{align}
    &A = \frac{J_1-J_2}{2}, \\
    &\varepsilon_0 = 3 \frac{(J_1 + J_2)}{2} - \frac{2}{3}D
\end{align}
for some non-negative values $J_{1}>J_2$ of exchange interaction strength.

\subsection{Bloch-sphere representation of stochastic resonance condition}
\label{sm:bloch-sphere-argument}
\begin{figure*}
    \centering
    \includegraphics[width=0.98\textwidth]{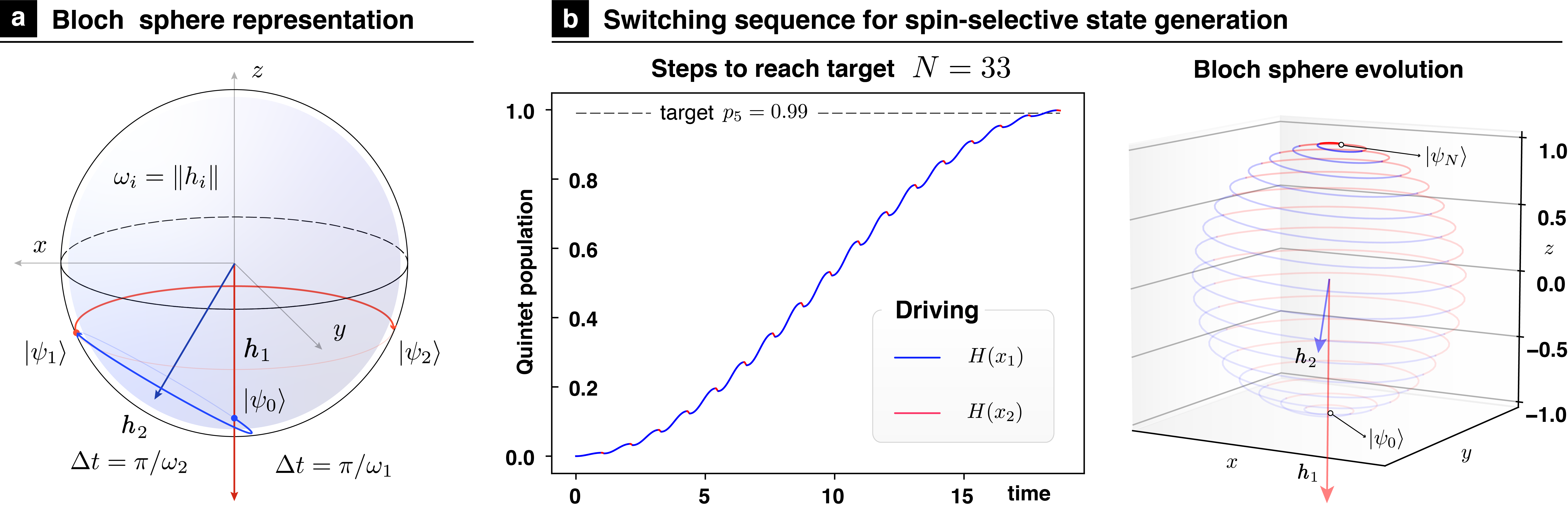}
    \caption{(\textit{Color online}) \textbf{\textsf{a} Spin mixing on the Bloch sphere.}---
    (\textit{Color online}) \textbf{\textsf{b} Switching sequence for spin-selective state preparation.}---(\textit{Left}) An initial singlet state $\ket{\psi}_0$ is driven within target population $p_5 = 0.99$ from the quintet state $\ket{\quintet}$ by means of a sequence of switching between $x_1$ and $x_2$, which varies the exchange strength $J(x_t)$. The target state $\ket{\psi_N}$ is reached after $N = 33$ steps, at time $t \approx 18 \:\omega_1/\pi$. (\textit{Right}) The same evolution on the Bloch sphere spanned by $\ket{\singlet}$ and $\ket{\quintet}$. Vectors $\bm{h}_i$ associated with Hamiltonians $H(x_i)$ not in scale.}
    \label{fig:sm_bloch_sphere}
\end{figure*}

\noindent
The results of Sec.~\ref{ss:nofield} and~\ref{ss:harmonic_conformation} can be interpreted by representing the spin mixing dynamics on the Bloch sphere spanned by the (orthogonal) singlet $\ket{\singlet}$ and quintet $\ket{\quintet}$ states that participate in the evolution, as shown in Fig.~\ref{fig:sm_bloch_sphere}~\textbf{\textsf{a}}~\cite{Bengtsson2006}. The Hamiltonian $H(x_1)$ and $H(x_2)$ are associated with vectors $\bm{h}_1$ and $\bm{h}_2$, respectively, via
\begin{equation}
    \label{eq:bloch_vector}
    H(x_i) \to \bm{h}_i = (-\Delta,0,-\varepsilon(x_i)).
\end{equation}
Under our strong-exchange regime assumption, and choosing $\varepsilon(x_1) > \varepsilon(x_2)$, we have $\varepsilon(x_1)\gg\Delta$, which implies that the singlet state $\ket{\singlet}$ is well approximated by one of the eigenstates of $\sigma_z$. Let us choose $\bm{r}_0 \leftrightarrow \ket{\psi_0}$ to be given by $-\hat{z}$, the \textit{south pole} of the Bloch sphere, without loss of generality, as shown in Fig.~\ref{fig:sm_bloch_sphere}~\textbf{\textsf{a}}. Accordingly, the quintet state $\ket{\quintet}\leftrightarrow \hat{z}$ corresponds to the \textit{north pole} of the Bloch sphere.

The dynamics of a state $\ket{\psi}_t \leftrightarrow \bm{r}_t$ on the Bloch sphere is given by the precession of its Bloch vector $\bm{r}_t$ around the vector $\bm{h}$ associated with some Hamiltonian $H$. The precession occurs at frequency $\omega = \|\bm{h}\|$, proportional to the Hilbert-Schmidt norm of the traceless Hamiltonian. 
In the strong-exchange regime, the two vectors $\bm{h}_1$ and $\bm{h}_2$ are almost parallel to each other (here their difference is exaggerated). In this case, the best chance to drive $\bm{r}_0$ towards $\hat{z}$ is to (1) let it reach the azimuth of its orbit (blue line) when driven by $\bm{h}_2$, and (2) reset its \textit{phase} on the $xy$-plane when driven by $\bm{h}_1$, to repeat the process and gain quintet character. This condition is met when $\Delta t = \pi/\omega_i$ for each step, where $i$ corresponds to the index of the driving Hamiltonian $H(x_i)$.
Indeed, we can exploit this strategy to design an elementary sequence of switching times $\{t_n\}_{n=1}^{N}$ that drives an initial singlet within target population of the quintet state, as shown in Fig.~\ref{fig:sm_bloch_sphere}~\textbf{\textsf{b}}. 

\subsection{Sampling Harmonic Conformational Dynamics}
\label{sm:sampling_harmonic_conformational_dynamics}

\noindent
In Sec.~\ref{ss:harmonic_conformation} we model $X$ as a classical harmonic mode with frequency $\omega$ that exchanges energy with the thermal bath at temperature $T$. The trajectories $x(t)$ of $X$ are modelled as harmonic oscillations interrupted by energy exchange events that vary energy (amplitude) and phase of the oscillation:
\begin{equation}
    \label{eq:perturbed_harmonic_oscillations}
    x_{i,i+1}(t) = A_{i} \cos(\omega t + \phi_i), \textrm{for } \: t\in[t_i,t_{i+1}).
\end{equation}

\noindent
The amplitude of the oscillations $A(E,\omega) = A_0 \sqrt{E/\hbar\omega}$ is normalised such that $A(E_0,\omega) = 1$ for $E_0$ being the energy of the mode at the beginning of the evolution, by setting $A_0 = \sqrt{\hbar\omega/E_0}$.
Energy exchange event vary the energy of the oscillator, $E\to E' = E+\Delta E$ at a rate given by an Ohmic spectral density $\gamma(\Delta E, T)$
\begin{equation}
    \label{eq:spectral_density}
    \gamma(\Delta E, T) = \gamma_0 \frac{\Delta E}{\Lambda} \frac{\exp(-|\Delta E|/E_c)}{1-\exp(-\Delta E/k_B T)},
\end{equation}
where $\Lambda, E_c = k_B T$ to ensure $\gamma(0,T) = \gamma_0$. 

The trajectories are numerically obtained by sampling the energy exchange events: First, a uniformly distributed random number $u\in(0,1]$ is generated to pick the energy exchange event $\Delta E_i$ according to the probability distribution $p(\Delta E) = \gamma(\delta E, T)/Q$ with $Q = \int ds \gamma(s, T)$. Then, the time interval $\Delta t$ after which the event occurs is picked using $\Delta t_i = \log(1/u)/Q$. A new phase $\phi_{i+1}$ for the oscillator is pick randomly from the uniform distribution over $[0,2\pi)$. The energy, amplitude and phase of the oscillator are updated according to $E_i\to E_{i+1} = E_i + \Delta E$, $A_i\to A_{i+1} = A(E_{i+1},\omega)$ and $\phi_i\to\phi_{i+1}$, at time is set to $t_{i+1} = t_i + \Delta t_i$. During the time interval $\Delta t_i$ the oscillator evolves according to Eq.~\eqref{eq:perturbed_harmonic_oscillations}. 
The exchange strength $J$ is linear on the conformational configuration $x$ as
\begin{equation}
    \label{eq:exhcange_harmonic}
    J(x) = \frac{J_\mathrm{max} - J_\mathrm{min}}{2} x + \frac{J_\mathrm{max} + J_\mathrm{min}}{2},
\end{equation}
where $J_\mathrm{min(max)} = \mathrm{min (max)}\{J_1,J_2\}$. Each trajectory is initialised such that $x_0 = 1$ so that $J(x_0) = J_\mathrm{min}$.

\end{document}